\documentclass[aps,showpacs,amsmath,amssymb,nofootinbib]{revtex4}
\usepackage{graphicx}
\usepackage{amssymb}
\newcommand{\mb}[1]{\mbox{\scriptsize #1}}
\newcommand{\sigmac}{\bar{q}\tau^{i}q}
\newcommand{\pic}{\bar{q}i\gamma_5\tau^{i}q}

\newcommand{\rhom}{\rho_{\mb{MRE}}}
\newcommand{\mre}{\int_{\mb{MRE}}}

\begin{document}
\title{Colour-superconducting strangelets in the Nambu--Jona-Lasinio model}
\author{O. Kiriyama}
\email{kiriyama@th.physik.uni-frankfurt.de}
\affiliation{Institut f\"ur Theoretische Physik, 
J.W. Goethe-Universit\"at, D-60439 Frankfurt am Main, Germany}

\date{\today}

\begin{abstract}
The two-flavour colour-superconducting (2SC) phase 
in small strangelets is studied. 
In order to describe the 2SC phase 
we use the three-flavour Nambu--Jona-Lasinio model. 
We explicitly take into account finite size effects 
by making use of the approximation for the density of states 
in spherical cavities called multiple reflection expansion (MRE). 
The thermodynamic potential for the 2SC strangelets is derived 
in the mean-field approximation with the help of the MRE. 
We found that 2SC phase survives in small strangelets 
with a sizable gap. Consequences for the 2SC phases are also discussed.  
\end{abstract}
\maketitle

\section{Introduction}
It is widely accepted that sufficiently cold and dense quark matter 
is a colour superconductor \cite{csc,revcsc}. 
At asymptotically high densities the perturbative one-gluon exchange 
interaction, directly based on the first principle of QCD, 
has been used to clarify the properties of colour superconductivity. 
At moderate densities, on the other hand, the studies of 
QCD-motivated effective theories have revealed rich phase structures. 

The most likely place for colour superconductivity is 
in the interior of compact stars. 
Therefore, the phases of quark matter with appropriate conditions 
for the interior of compact stars 
(i.e. neutrality conditions and $\beta$-equilibrium conditions) 
has attracted a great deal of interest. 
The most striking feature of neutral and $\beta$-equilibrated 
colour-superconducting phases is the appearance 
of gapless colour superconductivity. 
In the case of a large strange quark mass, 
it has been revealed that the phase called gapless 
two-flavour colour-superconducting phase (g2SC) \cite{g2SCa,g2SCb} 
could appear as a ground state. 
Similar analyses have been done 
for the case of three flavours 
and clarified that the gapless colour-flavour-locked (CFL) 
phase could also appear in neutral and $\beta$-equilibrated strange 
quark matter \cite{gCFLa,gCFLb}. 
(Note that, however, the gapless color superconductors develop 
a chromomagnetic instability \cite{ci2,ci3}.)

There exists other candidates for a colour superconductor 
called strangelets that are small chunks of 
strange quark matter \cite{SQM,STRT}. 
Strangelets with baryon number $A \ll 10^7$ 
are free from (local) electric neutrality condition 
in contrast to strange quark matter in bulk. 
As a consequence, the phases of strangelets could be different from 
that of bulk strange quark matter. 
Instead, we need to take into account finite-size effects explicitly 
in the study of strangelets. 
The properties of strangelets have been extensively studied 
by using various models 
with finite-size effects. 
In particular, Madsen \cite{CFLS} has shown that 
colour-flavour-locked strangelets are significantly more stable 
than unpaired (normal) strangelets, 
although the overall energy scale is still an open question 
because the bag constant is a phenomenological input parameter. 
In previous preliminary work \cite{KH}, we investigated 
the behaviour of chiral symmetry in finite size quark droplets 
consisting of up and down quarks 
within the two-flavour Nambu--Jona-Lasinio (NJL) model \cite{NJ}. 
In the NJL model, by contrast, 
the bag constant is generated dynamically \cite{SQMNJL}. 
To take into account finite-size effects, 
we used the so-called multiple reflection expansion (MRE) 
\cite{MRE,FJ,BJ,JMc}. 
In the MRE framework, the finite-size effects are included 
in terms of the density of states. 
The MRE has been used to calculate the thermodynamic quantities 
of finite-size quark droplets and reproduced well 
the results of mode-filling calculations \cite{JMc}. 

In this paper, we focus on the quark core of strangelets 
with baryon number $A \leq 10^4$ 
which are small enough to neglect electrons. 
In order to describe colour-superconductivity we use 
the three-flavour NJL model 
and study the conventional two-flavour colour superconducting (2SC) phase. 
Finite-size effects are incorporated in the thermodynamic potential 
by making use of the MRE. 
We deal with 2SC strangelets embedded in the physical vacuum. 
We do not discuss absolute stability of 2SC strangelets 
(as compared to a gas of $^{56}$Fe), but 
how the 2SC gap behaves in strangelets 
and its effects on properties of strangelets.

This paper is organised as follows. 
In Sec. II we formulate the thermodynamic potential 
for spherical 2SC strangelets. 
We concentrate on 2SC strangelets embedded in the physical vacuum 
and derive a set of coupled equations so as to find stable 
(i.e. pressure balanced) 2SC strangelets. 
In Sec. III we restrict ourselves to zero temperature 
and present numerical results. Section IV is devoted to 
summary and discussions.
 
\section{Thermodynamic potential for 2SC strangelets}
To describe the 2SC phase 
we use the $\mbox{U}(3)_L \times \mbox{U}(3)_R$ symmetric NJL model. 
We assume that the strange quark is sufficiently heavy and 
does not take part in Cooper pairing. 
The Lagrangian density is given by
\begin{eqnarray}
{\cal L}&=&\bar{q}(i\gamma^{\mu}\partial_{\mu}-\hat{m})q
+G_S\sum_{i=0}^{8}\left[(\sigmac)^2+(\pic)^2\right]\nonumber\\
&&+G_D\left(\bar{q}i\gamma_5\tau^{2}\lambda^{2}C\bar{q}^T\right)
\left(q Ci\gamma_5\tau^{2}\lambda^{2}q\right)
\label{eqn:lag}
\end{eqnarray}
where $q$ denotes a quark field with three flavours ($N_f=3$) 
and three colours ($N_c=3$), $\hat{m}=\mbox{diag}(0,0,m_s)$ 
is the quark mass matrix and $C$ is the charge conjugation matrix, 
defined by $C^{-1}\gamma_{\mu}C=-\gamma_{\mu}^T$ and $C^T=-C$. 
The coupling constants $G_S$ and $G_D$ have the dimension of 
$(\mbox{mass})^{-2}$. The Gell-Mann matrices 
$\tau^{i}$ $(i=1,\cdots,8)$ with $\tau^0=\sqrt{2/3}\openone_f$ act 
in the flavour space and 
$\lambda^2$ is the antisymmetric generator of $\mbox{SU}(3)_c$. 

We choose the model parameters 
(the ultraviolet cutoff and the coupling constant) as follows: 
$\Lambda_{\mb{UV}}=0.6$ GeV and $G_S=6.42~\mbox{GeV}^{-2}$. 
The diquark coupling constant is set to $G_D=3G_S/4$, 
unless stated otherwise. 
The value of the strange quark mass 
shall be explicitly indicated in the following figures. 

In this paper we work in the mean-field approximation. 
In the mean-field approximation, we obtain 
the following Hamiltonian density:
\begin{eqnarray}
{\cal H}_{\mb{MFA}}&=&
\bar{q}(i\vec{\gamma}\cdot\vec{\nabla}+\hat{m})q
+\frac{1}{2}
\left(\bar{q}i\gamma_5\tau^2\lambda^2C\bar{q}^T+\mbox{h.c.}\right)\nonumber\\
&&+\frac{\Delta^2}{4G_D},\label{eqn:mfah}
\end{eqnarray}
where $\Delta=2G_D\langle\bar{q}i\gamma_5\tau^2\lambda^2C\bar{q}^T\rangle$ 
is the 2SC gap parameter. 
Using ${\cal H}_{\mb{MFA}}$, we can straightforwardly calculate 
the thermodynamic potential of bulk quark matter. 

In this work, however, we need to incorporate 
finite-size effects into the thermodynamic potential. 
To this end, we use the density of states 
derived from the MRE \cite{MRE,FJ,BJ,JMc}. 
In the MRE framework, the density of states 
for a spherical system is written as $k^2\rhom/(2\pi^2)$, 
where $\rhom=\rhom(k,m,R)$ is given by
\begin{eqnarray}
\rhom=1+\frac{6\pi^2}{kR}f_S\left(\frac{k}{m}\right)+
\frac{12\pi^2}{(kR)^2}f_C\left(\frac{k}{m}\right)+\cdots.\label{eqn:dos}
\end{eqnarray}
Here $m$ denotes the (Dirac) mass of quarks 
and $R$ is the radius of the sphere. 
The functions $f_S(k/m)$ and $f_C(k/m)$ represent 
the surface and curvature contributions to 
the fermionic density of states in the spherical cavity, respectively. 
The ellipsis corresponds to higher order terms in $1/R$, 
which are neglected throughout. 
The functional forms of $f_S(k/m)$ and $f_C(k/m)$ are given by
\begin{subequations}
\begin{eqnarray}
f_S\left(\frac{k}{m}\right)=\frac{-1}{8\pi}
\left(1-\frac{2}{\pi}\arctan\frac{k}{m}\right),\label{eqn:fs}
\end{eqnarray}
\begin{eqnarray}
f_C\left(\frac{k}{m}\right)=\frac{1}{12\pi^2}
\left[1-\frac{3k}{2m}\left(\frac{\pi}{2}-\arctan\frac{k}{m}\right)\right]
\label{eqn:fc}.
\end{eqnarray}
\end{subequations}
It should be noted that the functional form of $f_C$ 
for an arbitrary quark mass has not been derived 
within the MRE framework. The functional form of 
Eq. ($\ref{eqn:fc}$) is the ansatz by Madsen \cite{JMc}.
Note also that Eqs. (\ref{eqn:fs}) and (\ref{eqn:fc}) have 
the following massless limits:
\begin{eqnarray}
\lim_{m \to 0}f_S(k/m)=0~,~~\lim_{m \to 0}f_C(k/m)=-1/(24\pi^2).
\end{eqnarray}
Figure 1 shows the MRE density of states. 
One can see that the finite-size effects reduce the density of states 
and $\rhom$ becomes negative at small momenta. 
To avoid the unphysical negative density of states 
we shall introduce an infrared cutoff 
$\Lambda_{\mb{IR}}$ in momentum space. 
For the case of $m \neq 0$, we numerically solve 
the equation $\rhom=0$ with respect to $k$ 
and use the larger one as $\Lambda_{\mb{IR}}$. 
On the other hand, for the case of $m=0$, 
the MRE density of states takes the form:
\begin{eqnarray}
\rhom(k,R)=1-\frac{1}{2(kR)^2}.
\end{eqnarray}
Then, we obtain $\Lambda_{\mb{IR}}=\sqrt{2}/(2R)$. 
We have confirmed that $\Lambda_{\mb{IR}}$ behaves like 
$\Lambda_{\mb{IR}} \propto 1/R$ in both cases. 
\begin{figure}
\includegraphics[scale=0.55]{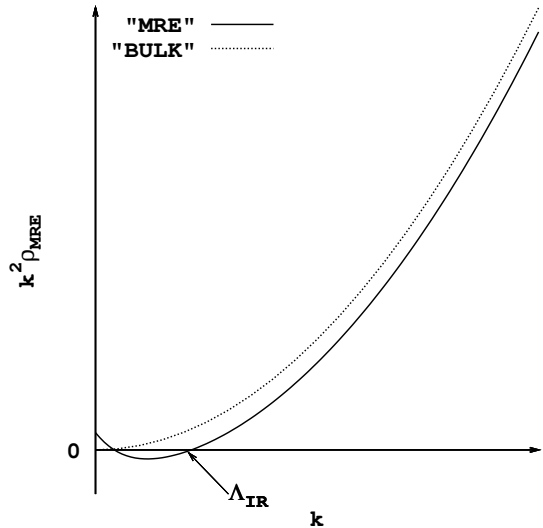}
\caption{The density of states $k^2\rhom$ for $m \neq 0$. 
We numerically solve the equation $\rhom=0$ 
and adopt the right-most root as infrared cutoff $\Lambda_{\mb{IR}}$.}
\end{figure}

Using the density of states (\ref{eqn:dos}), 
we express the effective potential (per unit volume) 
of the spherical strangelets as follows
\begin{widetext}
\begin{eqnarray}
\Omega_{\mb{MRE}}(\Delta;\mu,T)&=&\frac{\Delta^2}{4G_D}
-2\mre\left\{\epsilon_{-}+\epsilon_{+}
+2T\ln(1+e^{-\beta\epsilon_{-}})
(1+e^{-\beta\epsilon_{+}})\right\}\nonumber\\
&&-4\mre\left\{\mbox{sgn}(\epsilon_{-})E_{-}+E_{+}
+2T\ln(1+e^{-\beta\mbox{sgn}(\epsilon_{-})E_{-}})
(1+e^{-\beta E_{+}})\right\}\nonumber\\
&&-3\mre\left\{\varepsilon_{-}+\varepsilon_{+}
+2T\ln(1+e^{-\beta\varepsilon_{-}})
(1+e^{-\beta\varepsilon_{+}})\right\},
\end{eqnarray}
where $\epsilon_{\pm}=k\pm\mu$ are the quasiparticle energies 
of the unpaired (blue up and blue down) quarks, 
$E_{\pm}=\sqrt{(k\pm\mu)^2+\Delta^2}$ are those 
of the gapped (red up, red down, green up, and green down) quarks, 
and $\varepsilon_{\pm}=\sqrt{k^2+m_s^2}\pm\mu$ are those of strange quarks.
We have also introduced the sign function $\mbox{sgn}(x)$, 
$\mbox{sgn}(x)=\pm 1$ for $x>0$ or $x<0$, 
and the following short-hand notation,
\begin{eqnarray}
\mre=\int_{\Lambda_{\mb{IR}}}^{\Lambda_{\mb{UV}}}
\frac{k^2dk}{2\pi^2}\rhom.
\end{eqnarray}
Notice that we used a common chemical potential $\mu$ for all nine quarks. 
In general, the matrix of quark chemical potentials for 
three-flavour quark systems can be written as
\begin{eqnarray}
\mu_{ij,\alpha\beta}=
(\mu\delta_{ij}-\mu_{e}Q_{ij})\delta_{\alpha\beta}
+\mu_{3}\delta_{ij}(T_3)_{\alpha\beta}
+\mu_{8}\delta_{ij}(T_8)_{\alpha\beta},
\end{eqnarray}
where $Q$, $T_3$ and $T_8$ are generators of $\mbox{U}(1)_{em}$, 
$\mbox{U}(1)_3$ and $\mbox{U}(1)_8$, respectively. 
The indices $i,j$ and $a,b$ refer to flavour and colour, respectively. 
Hence, we should introduce three quark chemical potentials 
$\mu_e$, $\mu_3$ and $\mu_8$ in the study of strange quark matter in bulk. 
However, we can neglect the (local) electric charge neutrality 
and set $\mu_e=0$ in the study of small strangelets. 
We also neglect $\mu_3$ and $\mu_8$, for simplicity. 

Now we derive the gap equation with the MRE. 
For computations in a finite system, 
we temporarily choose a fixed radius $R$. 
Then, the extremum condition of $\Omega_{\mb{MRE}}$ 
with respect to $\Delta$ reads
\begin{eqnarray}
\Delta=8G_D\mre
\left\{\frac{\Delta}{E_{-}}\left[1-2N_F(E_{-})\right]
+\frac{\Delta}{E_{+}}\left[1-2N_F(E_{+})\right]\right\}.\label{eqn:sde}
\end{eqnarray}
This equation has a trivial solution ($\Delta=0$) 
as well as a nontrivial solution ($\Delta \neq 0$). 
The former corresponds to unpaired (normal) strangelets 
and the latter to colour-superconducting strangelets. 

In order to look at stable (pressure balanced) strangelets, 
we employ the usual pressure balance relation between 
(the inside of) a strangelet and the outer physical vacuum. 
The inside pressure $P$ can be defined by
\begin{eqnarray}
P=P_{\mb{MRE}}-B,\label{eqn:pb}
\end{eqnarray}
where $P_{\mb{MRE}}=-\Omega_{\mb{MRE}}$ and 
\begin{eqnarray}
B&=&
12\int\frac{d^3k}{(2\pi)^3}\sqrt{k^2+M_u^2}-\frac{M_u^2}{4G_S}\nonumber\\
&&+6\int\frac{d^3k}{(2\pi)^3}\sqrt{k^2+M_s^2}-\frac{(M_s-m_s)^2}{8G_S},
\end{eqnarray}
where $M_u$ and $M_s$ are the dynamically generated quark masses 
and they are the nontrivial solutions to the following equations,
\begin{subequations}
\begin{eqnarray}
M_u=24G_S\int\frac{d^3k}{(2\pi)^3}\frac{M_u}{\sqrt{k^2+M_u^2}},
\end{eqnarray}
\begin{eqnarray}
M_s&=&m_s+24G_S\int\frac{d^3k}{(2\pi)^3}\frac{M_s}{\sqrt{k^2+M_s^2}}.
\end{eqnarray}
\end{subequations}
Of course, these equations can be derived 
from the extremum condition of the effective potential 
at $T=\mu=0$. 
In Eq. (\ref{eqn:pb}), we have introduced 
the vacuum pressure (or bag constant) $B$ 
so as to measure the pressure relative to the outer vacuum. 
We emphasise that we do not need to introduce the bag constant by hand. 
In the NJL model, the bag constant is generated dynamically.

We first solve the gap equation [Eq. (\ref{eqn:sde})] 
and pressure balance relation $P=0$ [Eq. (\ref{eqn:pb})] self-consistently. 
Then, the baryon number $A$ of the stable strangelet is computed 
by using the following relation
\begin{eqnarray}
A=Vn_B,
\end{eqnarray}
where $V=4\pi R^3/3$ is the volume of the spherical strangelet and 
$n_B$ is the baryon number density. 
The baryon number density, 
which is one third of the quark number density $n_q$, 
is obtained by taking the partial derivative 
of $\Omega_{\mb{MRE}}$:
\begin{eqnarray}
n_B&=&n_q/3,\\
n_q&=&-\frac{\partial\Omega_{\mb{MRE}}}{\partial\mu}\nonumber\\
&=&4\mre\left\{N_F(\epsilon_{-})-N_F(\epsilon_{+})
-\frac{\epsilon_{-}}{E_{-}}\left[1-2N_F(E_{-})\right]
+\frac{\epsilon_{+}}{E_{+}}\left[1-2N_F(E_{+})\right]
+\frac{3}{2}\left[N_F(\varepsilon_{-})-N_F(\varepsilon_{+})\right]\right\}.
\end{eqnarray}
\end{widetext}
where $N_F(x)=1/(e^{\beta x}+1)$ is the Fermi distribution function. 

Having solved the above three equations self-consistently, 
we can find stable colour-superconducting strangelets 
(i.e. $A$ dependence of $\Delta$, $\mu$ and $R$). 
We can also compute other thermodynamic quantities. 
The most important, perhaps, is the energy per baryon number $E/A$ 
of strangelets:
\begin{eqnarray}
\frac{E}{A}\bigg{|}_{P=0}=\frac{{\cal E}}{n_B}\bigg{|}_{P=0},
\end{eqnarray}
where the energy density ${\cal E}$ is given by
\begin{eqnarray}
{\cal E}=\Omega_{\mb{MRE}}+\sum_{f=u,d,s}\mu_{f}n_{f}+B.
\end{eqnarray}
Here $n_f$ is the number density of quark flavour $f$ 
that satisfy the relation $\sum_{f}n_f=n_q$. 
By virtue of the common chemical potential $\mu$ 
and the pressure balance relation, 
one can see that the following relation holds:
\begin{eqnarray}
\frac{E}{A}\bigg{|}_{P=0}=3\mu.\label{eqn:epb}
\end{eqnarray}

\section{Numerical results}
In this section we focus on strangelets at zero temperature and 
present numerical results. 
Before proceeding to the results, we give a brief survey 
of the finite-size effects. 
Suppose a finite-size quark droplet embedded in the physical vacuum. 
It is generally known that the finite-size quark droplet is 
energetically disfavoured as compared to quark matter in bulk. 
In other words, the finite-size effects increase the 
quark chemical potential. 
Hence, our results show a size dependence (i.e. baryon number dependence). 

Let us start with the baryon number dependence of the 2SC gap (see Fig. 2). 
We present the results for the cases 
$m_s=0$ and $m_s=0.12$ GeV, for comparison. 
It is clear that, for relatively large baryon numbers ($A \agt 100$), 
the gap remains approximately constant. 
The result indicates that finite-size effects 
becomes less important for large baryon numbers (i.e. large radii), 
as they should. 
The difference in the size of the gap between 
the case $m_s=0$ and $m_s=0.12$ GeV is the result of 
the difference in the chemical potentials of the gapped quarks. 
At fixed $A$, the effect of $m_s$ is to 
increase the chemical potentials of up and down quarks. 
Consequently, the gap grows with $m_s$.

In contrast, the curves behave differently 
at small baryon numbers ($A \alt 100$). 
In the case of $m_s=0.12$ GeV, as $A$ is decreased 
the gap decreases. The decrease of the gap is caused by the fact that 
the chemical potentials of gapped quarks are increased 
by finite-size effects and, then, 
they are close to the UV cutoff (see Fig. 3). 

In the case of $m_s=0$, 
the behaviour of the gap at small baryon numbers 
is also due to the finite-size effects. 
However, the chemical potentials of the gapped quarks 
are not close to the UV cutoff (Fig. 3). 
As $A$ is decreased, the chemical potentials of gapped quarks 
are increased by the finite-size effects. 
Hence the gap grows with decreasing $A$. 
As $A$ is further decreased the gap reaches a peak and, finally, drops. 
The decrease of the gap at very small baryon numbers ($A\alt 10$) 
arises from the fact that 
the density of states near the Fermi surface is decreased 
by the IR cutoff.

Figure 4 shows energy per baryon number $E/A$ 
of 2SC strangelets as a function of $A$. 
The strange quark masses are set to $m_s=0.12$ GeV. 
We also present the results 
for the cases of strong diquark coupling ($G_D=G_S$) 
and unpaired strangelets. 
It appears that 2SC strangelets are more stable 
than unpaired strangelets and the strong diquark coupling 
significantly stabilises 2SC strangelets. 
The results can be easily understood as follows. 
In the 2SC phase, the pairing energy contribution 
to the thermodynamic potential is roughly given by (volume term only)
\begin{eqnarray}
\Omega_{p}^{\mb{(2SC)}}=-\frac{\Delta^2\mu^2}{\pi^2},
\end{eqnarray}
that is the product of the pairing energy 
and the number density of gapped quarks near the Fermi surface \cite{AR}. 
Therefore, it is quite reasonable that 
2SC strangelets are more stable than unpaired strangelets. 
It is also natural that the strong diquark coupling 
(i.e. large paring energy) makes 2SC strangelets energetically more favoured. 
For small $A$ the energy per baryon number increases dramatically 
because of the growth of the chemical potential 
by the finite-size effects. 

To examine $E/A$ of various phases we compute 
$E/A$ for unpaired, 2SC, and CFL strangelets 
as a function of $A$. (See Fig. 5. The curve for CFL strangelets 
is taken from Ref. \cite{Kiriyama}). 
The diquark coupling and the strange quark mass 
are taken to be $G_D=3G_S/4$ and $m_s=0$, respectively. 
The pairing energy in the CFL phase takes the following form \cite{AR}:
\begin{eqnarray}
\Omega_{p}^{\mb{(CFL)}}=-\frac{3\Delta^2\mu^2}{\pi^2}.
\end{eqnarray}
Hence the pairing energy contribution to the thermodynamic potential 
in the CFL phase is larger than that in the 2SC phase 
because all nine quarks participate in Cooper pairing. 
As one would expect, CFL strangelets are more stable than 2SC strangelets. 
In general, however, we need to include nonvanishing $m_s$ 
and study the competition of 2SC and CFL phases in detail.

It is interesting to look at the strangeness fraction of 
2SC strangelets in the light of their production in laboratories. 
The strangeness fraction $f_s$ is given by
\begin{eqnarray}
f_s=\frac{n_s}{n_u+n_d+n_s}.
\end{eqnarray}
The fraction for unpaired strangelets should satisfy 
$f_s=1/3$ at $m_s=0$. 
Figure 6 shows the strangeness fraction of 2SC strangelets 
for the cases of $m_s=0$, $m_s=0.12$ GeV, and $m_s=0.15$ GeV. 
The diquark coupling is taken to be $G_D=3G_S/4$. 
The fraction deviates from $f_s=1/3$ even in the case of $m_s=0$. 
This deviation can be understood as follows. 
The number density of an unpaired quark is given by
\begin{eqnarray}
n_0=2\mre\theta(\mu-k),
\end{eqnarray}
where $\theta(\mu-k)$ is the occupation number 
for a noninteracting massless quark. 
On the other hand, the number density of a gapped quark is 
given by 
\begin{eqnarray}
n_{\Delta}=2\mre\frac{E_{-}+\mu-k}{2E_{-}},
\end{eqnarray}
where $(E_{-}+\mu-k)/(2E_{-})$ is 
the occupation number of a gapped quark. 
One can easily check that the relation $n_{\Delta}>n_{0}$ holds. 
Hence, the number of gapped quarks increases in the 2SC phase 
and the relation $f_s<1/3$ holds even in the case of $m_s=0$. 
For the cases of massive strange quark, 
the fraction decreases further because 
the Fermi momentum of strange quarks, $k_F=\sqrt{\mu^2-m_s^2}$, 
decreases due to the mass gap. 
One may argue that $f_s$ is reduced further 
if one takes into account the dynamically generated mass of the strange quark. 
However, it has been shown that 
the finite-size effects enhance the restoration of chiral symmetry \cite{KH}. 
Therefore, at present, the effect of the dynamical strange quark mass 
is an open question. 
The strangeness fraction in CFL strangelets is, presumably, larger 
than that of 2SC strangelets, because CFL pairing 
enforces the equal Fermi momenta for all quarks. 
The complete calculation including dynamical quark masses is 
an interesting problem to be investigated in future work.

Figure 7 shows the densities achieved in 2SC strangelets 
as a function of $A$. 2SC strangelets with massive ($m_s=0.12$ GeV) 
strange quarks have higher densities 
than that with massless strange quarks. 
This is understood by considering 
that nonzero $m_s$ gives a contribution to 
the pressure. 
The leading-order contribution from nonvanishing $m_s$ 
is roughly given by (volume term only)
\begin{eqnarray}
\Omega_s=\frac{3m_s^2\mu^2}{4\pi^2}.
\end{eqnarray}
Comparing the cases of $m_s=0$ and $m_s=0.12$ GeV, 
we have confirmed that $\Omega_{s}$ overpowers the small difference of 
$\Omega_{p}^{\mb{(2SC)}}$. 
Then 2SC strangelets with massive strange quarks require 
larger chemical potentials to maintain the pressure balance 
and, therefore, have larger density than that with massless strange quarks. 

Inversely, the radii of strangelets behave as shown in Fig. 8. 
The size of 2SC strangelets 
at very small baryon numbers ($A \simeq 10$) 
is around 1--2 fm. 
The size is comparable with the coherence length of the Cooper pair 
$\xi=1/(\pi\Delta)$. 
Then, it is uncertain whether the Cooper pair exists 
in such a small system or not. 
In these regions, mode-filling calculations may be available \cite{ABMW}. 

\begin{figure}
\includegraphics[scale=0.55]{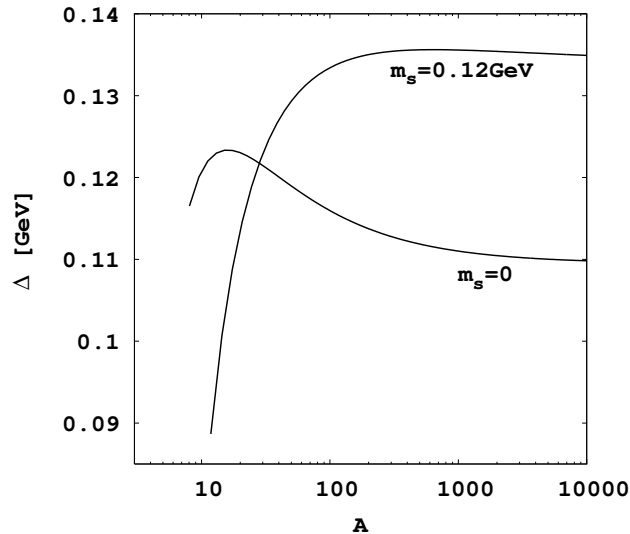}
\caption{The 2SC gap $\Delta$ as a function of baryon number $A$ 
for the cases of $m_s=0$ and $m_s=0.12$ GeV. 
The diquark coupling is taken to be $G_D=3G_S/4$.}
\end{figure}

\begin{figure}
\includegraphics[scale=0.55]{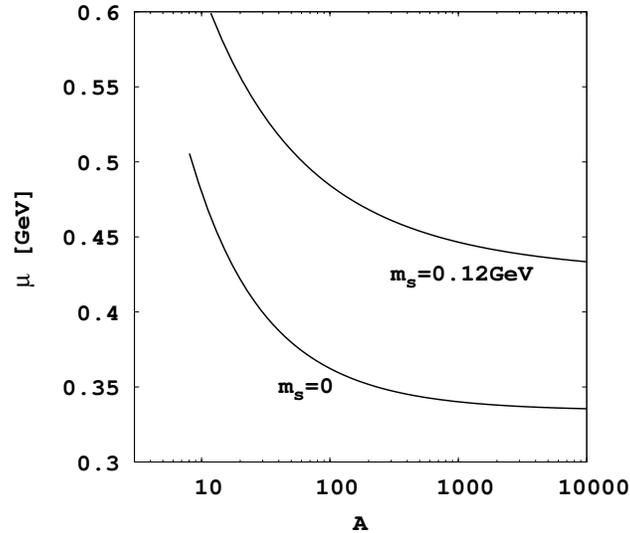}
\caption{The quark chemical potentials $\mu$ of the 
pressure balanced 2SC strangelets as a function of baryon number $A$. 
Recall that the UV cutoff is taken to be $\Lambda_{\mb{UV}}=0.6$ GeV.}
\end{figure}

\begin{figure}
\includegraphics[scale=0.55]{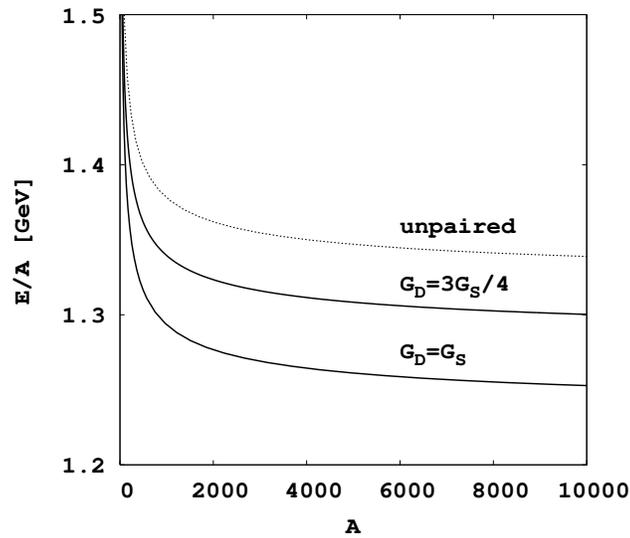}
\caption{\label{fig:Fig4}
The energy per baryon number $E/A$ as a function of $A$ 
for 2SC strangelets (solid lines) and unpaired strangelets (dotted line).}
\end{figure}

\begin{figure}
\includegraphics[scale=0.55]{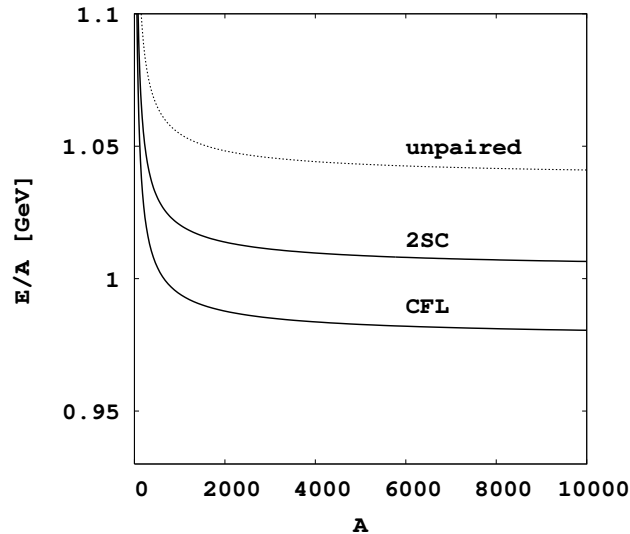}
\caption{\label{fig:Fig5}
The energy per baryon number $E/A$ as a function of $A$ 
for CFL strangelets (bottom), 2SC strangelets (middle) 
and unpaired strangelets (top). 
The diquark coupling and the quark mass are set to 
$G_D=3G_S/4$ and $m_s=0$, respectively.}
\end{figure}

\begin{figure}
\includegraphics[scale=0.55]{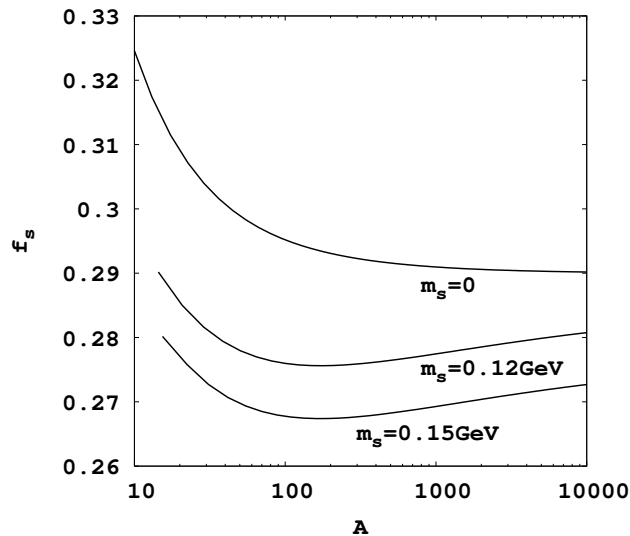}
\caption{The strangeness fraction $f_s$ of 2SC strangelets 
as a function of baryon number $A$. 
The diquark coupling is set to $G_D=3G_S/4$. 
The strange quark masses are taken to be $m_s=0,0.12,0.15$ GeV 
(from top to bottom).}
\end{figure}

\begin{figure}
\includegraphics[scale=0.55]{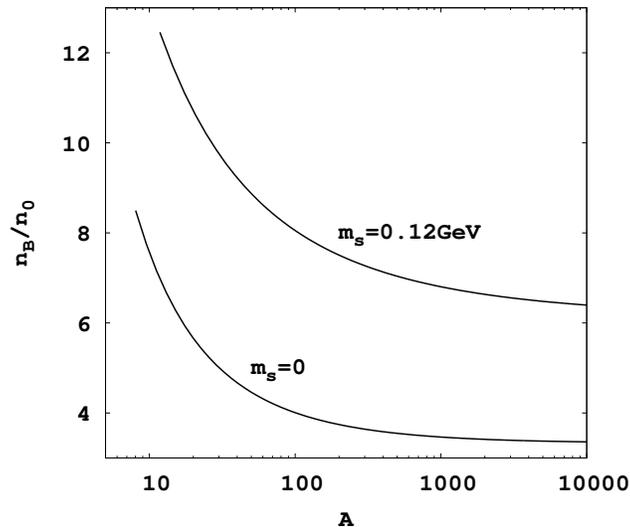}
\caption{The baryon number density $n_B$ in 2SC strangelets divided by 
the normal density $n_0=0.17\mbox{fm}^{-3}$.}
\end{figure}

\begin{figure}
\includegraphics[scale=0.55]{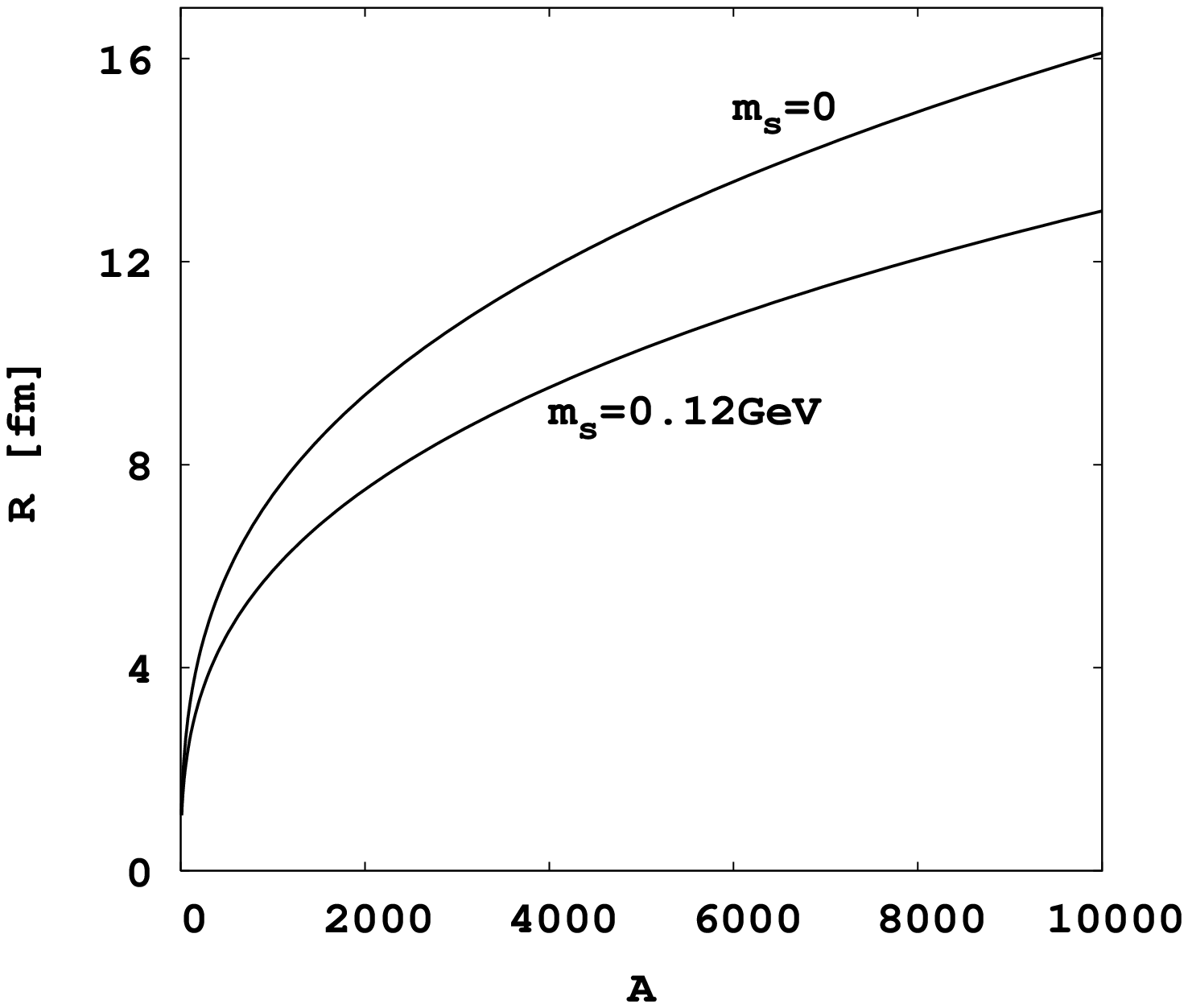}
\caption{The radius $R$ of 2SC strangelets 
as a function of baryon number $A$ 
for $m_s=0,0.12$ GeV. Both curves behave like $R \propto A^{1/3}$.}
\end{figure}

\section{Summary and discussions} 
In summary, we studied the properties of 
two-flavour colour-superconducting strangelets. 
We used the Nambu--Jona-Lasinio model to describe 
colour superconductivity and the multiple reflection expansion to 
take account of finite-size effects. 
We formulated the thermodynamic potential for 2SC strangelets 
including finite-size effects in terms of the MRE density of states. 
We then solved the set of coupled equations; 
the gap equation, the pressure balance relation 
and the baryon number condition. 
We found that a sizable 2SC gap survives 
in small strangelets ($A \simeq 100$) 
though its behaviour at small baryon numbers 
depends on the strange quark mass. 
We also found that 2SC strangelets are 
more stable than unpaired strangelets due to the pairing energy. 
It should be noted here that 
2SC strangelets are not absolutely stable 
as compared to a gas of $^{56}$Fe. 
We did not examine the competition with other possible phases 
(e.g. colour-flavour-locked phase, chirally broken phase, and so on). 
We would like to make a comment on the sensitivity of the results 
to model parameters. 
Using several sets of parameters, 
we examined the sensitivity of our results. 
We found that the results do not change much 
with the choice of $\Lambda_{\mb{UV}}$ and $G_S$, 
as long as they are fixed by fitting physical quantities 
at $T=\mu=0$. 
Rather than $\Lambda_{\mb{UV}}$ and $G_S$, 
the diquark coupling has an effect on the results. 
A large diquark coupling yields a large 2SC gap and 
reduces the strangeness fraction. 
For instance, the strangeness fraction decreases to 
$f_s \simeq 0.25$ at $m_s=0.12$ GeV and $G_D/G_S=1.2$ 
due to the large 2SC gap ($\Delta \simeq 0.28$ GeV). 
However, such a strong coupling may induce the fluctuation of 
pairing field and, then, invalidate the present description of the 2SC phase.

The MRE has problems concerning its reliability. 
First, $\rhom$ should have terms proportional to 
$1/R^2$, $1/R^4$ and so on. They are dominant at small radii. 
Further, the MRE causes a negative density of states. 
Although we simply avoided the latter problem 
by introducing an infrared cutoff, 
these problems should be solved in the future. 
However, for relatively large systems ($A \agt 100$), 
these problems have rather minor effects 
and the results presented in this paper would hold.

Finally, we comment on the outlook for future studies. 
It is very interesting to study the competition with other phases, 
taking account of the realistic value of the strange quark mass. 
In particular, inclusion of the chirally broken phase 
and the colour-flavour-locked phase would affect the present analysis. 
It would be also interesting 
to examine colour-superconducting strangelets 
embedded in the vacuum at finite temperature and/or density. 

\begin{acknowledgments}
I would like to thank D. Rischke and I. Shovkovy 
for discussions and a critical reading of the manuscript. 
\end{acknowledgments}

\end{document}